\documentclass[aps,preprint,nofootinbib,floatfix]{revtex4-1}
\usepackage{graphicx,color}
\usepackage{amsmath,amssymb}
\usepackage{url}
\usepackage{epstopdf}
\newcommand{\lsim}{\mathrel{\mathop{\kern 0pt \rlap
  {\raise.2ex\hbox{$<$}}}
  \lower.9ex\hbox{\kern-.190em $\sim$}}}
\newcommand{\gsim}{\mathrel{\mathop{\kern 0pt \rlap
  {\raise.2ex\hbox{$>$}}}
  \lower.9ex\hbox{\kern-.190em $\sim$}}}

\begin{document}
\title{Gamma-ray Constraints on Effective Interactions\\
 of the Dark Matter}
\author{Kingman Cheung$^{1,2}$, Po-Yan Tseng$^{2}$ and Tzu-Chiang Yuan$^3$}

\affiliation{
$^1$Division of Quantum Phases \& Devices, School of Physics, 
Konkuk University, Seoul 143-701, Korea \\
$^2$Department of Physics, National Tsing Hua University, 
Hsinchu 300, Taiwan
\\
$^3$Institute of Physics, Academia Sinica, Nankang, Taipei 11529, Taiwan
}

\date{\today}

\begin{abstract}
  Using an effective interaction approach to describe the interactions
  between the dark matter particle and the light degrees of freedom of
  the standard model, we calculate the gamma-ray flux due to the
  annihilation of the dark matter into quarks, followed by
  fragmentation into neutral pions which subsequently decay into
  photons.  By comparison to the mid-latitude
  data released from the Fermi-LAT experiment,
  we obtain useful constraints on the size of the effective interactions 
  and they are found to be comparable to those deduced 
  from collider, gamma-ray line and anti-matter search experiments.
  However, the two operators induced by scalar and vector exchange 
  among fermionic dark matter and light quarks
  that contribute to spin-independent cross sections are constrained more stringently 
  by the recent XENON100 data.
\end{abstract}
\maketitle

\section{Introduction}
The presence of cold dark matter (CDM) in our Universe is now well established
by a number of observational experiments, especially the very precise 
measurement of the cosmic microwave background radiation
in the Wilkinson Microwave Anisotropy Probe (WMAP) experiment \cite{wmap}.
The measured value of the CDM relic density is
\begin{equation}
\label{wmap}
 \Omega_{\rm CDM}\, h^2 = 0.1126 \;\pm 0.0036 \;,
\end{equation}
where $h$ is the Hubble constant in units of $100$ km/Mpc/s.
Though the gravitation nature of the dark matter is 
commonly believed to be well established, 
its particle nature remains allure except that it is
nonbaryonic and to a high extent electrically neutral. 

One of the most appealing and natural CDM particle candidates is the
{\it weakly-interacting massive particle} (WIMP).  It is a coincidence
that if the dark matter (DM) $\chi$ is thermally produced  in the early Universe,
the required annihilation cross section is right at the order of
weak interaction.  The relation between the 
fractional relic density of $\chi$ 
relative to the critical density and
its thermal annihilation cross section can be given by the following 
simple formula \cite{hooper}
\begin{equation}
\label{rate}
\Omega_\chi h^2 \simeq \frac{ 0.1 \;{\rm pb} }{\langle \sigma v \rangle} \;,
\end{equation}
with $\langle \sigma v \rangle$ being the 
annihilation cross section of the dark matter around the time of freeze-out,
at which the annihilation rate could no longer catch up with the 
Hubble expansion rate of the Universe.
Assuming the measured $\Omega_{\rm CDM} h^2$ to be 
saturated by a single component WIMP, its annihilation cross section
should be about $1$ pb or $3\times 10^{-26}\;{\rm cm}^3 \, {\rm s}^{-1}$.  
This is exactly the size of the cross section that one expects from 
a weak interaction process,
which implies an appreciable
size of production rate of the WIMP at the Large Hadron Collider (LHC)
as well as the event rates for direct and indirect 
searches that reach the sensitivities  of dark matter experiments 
like XENON100 and Fermi-LAT respectively.
In general, production of dark matter at the LHC would give rise to
a large missing energy.  Thus, the anticipated signature in the 
final state is high-$p_T$ jets or leptons plus a large missing energy.
Note that there could be non-thermal sources for the dark matter, such 
as decay from exotic relics like moduli fields, cosmic strings, etc. In 
such cases, the annihilation cross section
 in Eq.~(\ref{rate}) can be larger than
the value quoted above.

There have been many proposed candidates for the dark matter. 
Without committing to any particular DM model so as 
to perform a model independent analysis,
we adopt an effective interaction
approach to describe the interactions of the dark matter particle with
the standard model (SM) particles 
\cite{cao,bai,tait,marv,Mack,tait-gamma,rosz,good,fan,ours,cty,celine,gauge-higgs}.
One simple realization of the effective interaction approach 
is that the dark matter particle exists in a 
hidden sector, which communicates to the SM sector via a heavy degree
of freedom in the connector sector. At energy scale well below this heavy 
mediator the interactions
can be conveniently described by a set of effective interactions.
The strength of each interaction depends on  the nature of the dark
matter particle and the mediator.
The most important set of interactions are among the fermionic dark matter 
$\chi$ and the light quarks $q$ described by the effective operators 
$(\bar \chi \Gamma \chi)(\bar q \Gamma^\prime q)$
where $\Gamma$  and $\Gamma^\prime$ are general Dirac matrices
contracted with appropriate Lorentz indices. We will discuss these and other 
operators in more details in the next section.

There have been some recent works on constraining the interactions 
at present and future collider experiments \cite{cao,bai,tait,marv},
using gamma-ray experiments \cite{Mack,tait-gamma,rosz,good,berg,HG} and
using anti-matter search experiments \cite{celine,cty,gauge-higgs}.
There was another work in which the dark matter couples
only to the top quark and corresponding predictions at direct and indirect
detection experiments as well as colliders were obtained \cite{ours}.
It was also shown in Ref.~\cite{Ciafaloni} that additional 
radiation of electroweak bosons in the final state can modify the energy 
spectrum,
especially at the lower end of the spectrum. 
Lifting the helicity suppression by 
radiating off an electroweak gauge boson from the external light fermion 
legs due to 
Majorana dark matter annihilation is emphasized
in Ref.~\cite{Ciafaloni,Weiler}.

In Ref.~\cite{tait-gamma}, monochromatic photon-line flux was 
calculated via a loop with quarks running in it and photons being 
attached to the internal quark line. Although the photon-line would be a 
smoking-gun signal to compare with the data, the rate is suppressed because of the
loop factor. On the other hand, photons can come from the decay of
neutral pions, which in turn come from the fragmentation of the quarks
in the annihilation of the dark matter. The chance that an energetic 
quark fragments into neutral pions is high and the branching ratio of
a neutral pion into two photons is 98.823\% \cite{pdg}.
Therefore, the amount
of photons coming from the quark fragmentation is much larger than 
those coming off a loop process.  Nevertheless, the spectrum of such
photons is continuous and in general have no structure, 
except for a cutoff due to the mass of the dark matter. 
In this work, we focus on the continuous gamma-ray flux spectrum
coming from the fragmentation of quarks into neutral pions, followed
by their decays into photons, in the annihilation of the dark matter.
Such annihilation of the dark matter will give rise to an additional source
of diffuse gamma-rays other than the known backgrounds.
If the experimental measurement is consistent with the known gamma-ray
background estimation, then one could use the data to constrain the
amount of gamma-ray flux coming from the dark matter annihilation, thus
constraining the effective interactions between the dark matter and
the quarks.

The data on the photon spectrum from the mid-latitude 
$(10^\circ < |b| < 20^\circ, \; 0^\circ < l < 360^\circ )$ 
\cite{fermi-lat}  recorded by
the Fermi-LAT indicated a continuous spectrum and mostly consistent
with the known backgrounds.  We can therefore use the data to constrain
on additional sources of gamma-ray, namely, the annihilation of the
dark matter into quarks, followed by fragmentation into neutral pions,
which further decay into photons.  The production of photons via
neutral pions is the dominant mechanism for gamma-rays.  There are also
other ways that the quarks from annihilation of dark matter can produce
gamma rays, such as bremsstrahlung off quark legs, synchrotron radiation,
or inverse Compton scattering on background photons, but these processes
are all $\alpha_{\rm em}$ suppressed relative to the fragmentation of the
quarks into neutral pions.  We focus on the fragmentation of the 
quarks coming from the annihilation of the dark matter as the signal 
in our analysis.  
We employ two approaches of obtaining the photon spectrum due to
fragmentation of light quarks. (i) We use the process 
$e^+ e^- \to q \bar q$ with initial radiations turned off in 
Pythia \cite{pythia} and extract the photon spectrum in the final state. 
The photon mainly comes from the decay of $\pi^0$, which are in turn
produced by fragmentation of light quarks, plus 
a very small fraction from the bremsstrahlung photon off the quark legs.  
(ii) We use the fragmentation function of
$q,\bar q, g$ into $\pi^0$ from the fitting of Ref.~\cite{kniehl}, then
convolute with the $dN/dE_\gamma (\pi^0 \to \gamma \gamma)$  to obtain
the photon spectrum of quarks into photon. We found that both approaches
give almost the same photon spectrum from light quarks. The resulting
limits using both approaches are also the same within numerical accuracy.
However, we have to use the second approach when we place limits on the
operators involving gluons, because we do not find an appropriate 
process in Pythia for extraction of $g\to \gamma$ fragmentation.

The choice of the mid-latitude data instead of the Galactic Center is simply
because the gamma-ray in this region is dominated by local sources and
we have clarity in understanding the background flux and point sources
within the mid-latitude. On the other hand, 
the Galactic center is supposed to have a number of known and known-unknown 
point sources, including a supermassive black hole near the Center, 
and perhaps some unknown sources too.  Given the purpose
of constraining the new DM interactions it is better to understand
clearly about the background in mid-latitude, rather than the 
larger flux from the Galactic Center.
The Galactic diffuse gamma rays originate primarily from the interactions
of high energy charged particles contained in cosmic rays with the nuclei
in the interstellar medium and the associated radiation fields of the 
charged particles, via a few mechanisms briefly described in Sec. III.
While most of them are well understood, the extra-galactic component
has a larger uncertainty.  We will choose a normalization such that
the total background diffuse gamma-ray flux is consistent with
the Fermi-LAT measurement of diffuse gamma-ray flux in the mid-latitude.
This approach is the same as the Fermi-LAT when they estimated the
extra-galactic diffuse component \cite{fermi-lat}.

The organization of the paper is as follows.  In the next section, we
describe the interactions between the dark matter particle and the SM
particles, in particular quarks and gluons.  
In Sec. III, we discuss various sources of diffuse gamma-ray flux that 
constitute the known background and 
calculate the gamma-ray flux due to the dark matter annihilation
using the effective interactions.
We compare with other constraints and conclude in Sec. IV.

\begin{table}[th!]
\caption{\small \label{table1}
The list of effective interactions between the dark matter and the light
degrees of freedom (quark or gluon). 
We have suppressed the color index
on the quark and gluon fields. These operators have also been
analyzed in Refs.~\cite{cao,tait,tait-gamma,cty}.  }
\begin{ruledtabular}
\begin{tabular}{lcr}
Operator & Coefficient & Velocity Scaling in $\langle \sigma v \rangle$ \\
\hline
\multicolumn{3}{l}{Dirac DM, (axial) vector/tensor exchange} \\
\hline
$O_1 = (\overline{\chi} \gamma^\mu \chi)\, (\bar q \gamma_\mu q)$ & 
                    $ \frac{C}{\Lambda^2}$ & $m_\chi^2$ \\
$O_2 = (\overline{\chi} \gamma^\mu \gamma^5\chi)\, (\bar q \gamma_\mu  q)$ & 
                    $ \frac{C}{\Lambda^2}$  & $m_\chi^2 v^2 $ \\
$O_3 = (\overline{\chi} \gamma^\mu \chi)\, (\bar q \gamma_\mu \gamma^5 q)$ & 
                     $\frac{C}{\Lambda^2}$ & $m_\chi^2$  \\
$O_4 = (\overline{\chi} \gamma^\mu \gamma^5 \chi)\, 
    (\bar q \gamma_\mu \gamma^5 q)$ &   $   \frac{C}{\Lambda^2}$   & $m_\chi^2 v^2 $ \\
$O_5 = (\overline{\chi} \sigma^{\mu\nu} \chi)\, (\bar q \sigma_{\mu\nu} q)$ & 
                     $\frac{C}{\Lambda^2}$  & $m_\chi^2$ \\
$O_6 = (\overline{\chi} \sigma^{\mu\nu} \gamma^5 \chi)\, 
  (\bar q \sigma_{\mu\nu} q)$ &  $\frac{C}{\Lambda^2} $   & $m_\chi^2$ \\
\hline
\multicolumn{3}{l}{Dirac DM, (pseudo) scalar exchange} \\
\hline
$O_7 = (\overline{\chi}  \chi)\, (\bar q  q)$ & $ \frac{C m_q }{\Lambda^3}$   & $m_q^2 m_\chi^2 v^2 $ \\
$O_8 = (\overline{\chi} \gamma^5  \chi)\, (\bar q  q)$ &  
                          $\frac{i  C m_q }{\Lambda^3}$    & $m_q^2 m_\chi^2$ \\
$O_9 = (\overline{\chi}  \chi)\, (\bar q \gamma^5 q)$& 
              $\frac{i C m_q }{\Lambda^3}$    & $m_q^2 m_\chi^2 v^2 $ \\
$O_{10} = (\overline{\chi} \gamma^5 \chi)\, (\bar q \gamma^5 q) $ &  
                        $ \frac{C m_q }{\Lambda^3}$    & $m_q^2 m_\chi^2$ \\
\hline
\multicolumn{3}{l}{Dirac DM, gluonic} \\
\hline
$O_{11} = (\overline{\chi} \chi)\, G_{\mu\nu} G^{\mu\nu}$ & 
                  $       \frac{C \alpha_s }{4 \Lambda^3} $    & $m_\chi^4 v^2 $ \\
$O_{12} = (\overline{\chi} \gamma^5 \chi)\, G_{\mu\nu} G^{\mu\nu} $ & 
                $         \frac{i C \alpha_s }{4 \Lambda^3} $    & $m_\chi^4$ \\
$O_{13} = (\overline{\chi} \chi)\, G_{\mu\nu} \tilde{G}^{\mu\nu}$ & 
                         $\frac{C \alpha_s }{4 \Lambda^3} $    & $m_\chi^4 v^2 $ \\
$O_{14} = (\overline{\chi} \gamma^5 \chi)\, G_{\mu\nu} \tilde{G}^{\mu\nu} $& 
                        $ \frac{i C \alpha_s }{4 \Lambda^3} $    & $m_\chi^4$ \\
\hline
\hline
\multicolumn{3}{l}{Complex Scalar DM, (axial) vector exchange} \\
\hline
$O_{15} = (\chi^\dagger \overleftrightarrow{\partial_\mu} \chi)\, 
        ( \bar q \gamma^\mu q ) $ & 
                     $    \frac{ C }{ \Lambda^2} $    & $m_\chi^2 v^2$ \\
$O_{16} = (\chi^\dagger \overleftrightarrow{\partial_\mu} \chi)\, 
  ( \bar q \gamma^\mu \gamma^5 q ) $ & 
                      $   \frac{ C }{ \Lambda^2} $    & $m_\chi^2 v^2$ \\
\hline
\multicolumn{3}{l}{Complex Scalar DM, (pseudo) scalar exchange} \\
\hline
$O_{17} = (\chi^\dagger \chi)\, ( \bar q  q ) $ & 
                       $  \frac{C m_q }{ \Lambda^2}$    & $m_q^2$ \\
$O_{18} = (\chi^\dagger \chi)\, ( \bar q  \gamma^5 q ) $ & 
                $         \frac{i C m_q }{ \Lambda^2} $    & $m_q^2$ \\
\hline
\multicolumn{3}{l}{Complex Scalar DM, gluonic} \\
\hline
$O_{19} = (\chi^\dagger \chi)\, G_{\mu\nu} {G}^{\mu\nu} $   & 
                       $  \frac{C \alpha_s }{4 \Lambda^2} $    & $m_\chi^2$ \\
$O_{20} = (\chi^\dagger \chi)\, G_{\mu\nu} \tilde{G}^{\mu\nu} $ & 
                      $   \frac{i C \alpha_s }{4 \Lambda^2} $    & $m_\chi^2$ \\
\end{tabular}
\end{ruledtabular}
\end{table}

\section{Effective Interactions}

For simplicity, we will assume there is only one component of dark matter 
denoted by $\chi$ and it is a standard model singlet.

The first set of operators we will be considering 
is for fermionic DM and
its effective interactions with light quarks via a 
(axial) vector- or tensor-type 
exchange are given by the following dimension 6 operators
\begin{equation}
\label{eff-q}
{\cal L}_{i=1-6} = \frac{C}{\Lambda_i^2} \,
   \left ( \overline{\chi} \Gamma \chi \right )\;
           \left ( \bar{q} \Gamma^\prime q \right ) \;,
\end{equation}
where  
$\Gamma , \Gamma^\prime = \gamma^\mu , \gamma^\mu \gamma^5, \sigma^{\mu\nu}$ or 
$\sigma^{\mu\nu} \gamma^5$
with 
$\sigma^{\mu\nu} \equiv i (\gamma^\mu \gamma^\nu -\gamma^\nu \gamma^\mu ) /2$.
$\Lambda_i$ is the heavy scale for the connector sector that has been 
integrated out and $C$ is an effective coupling constant of order $O(1)$.
It is understood that for Majorana fermion the vector and tensor 
structures of $\Gamma$ are absent.
Thus, for vector or tensor type interaction 
the fermion $\chi$ in Eq.(\ref{eff-q}) is understood to be Dirac. 
We will be focusing on Dirac fermionic DM in this work, but our results
are also applicable to Majorana dark matter.  
Note also that due to the following identity
\begin{equation}
\label{an-identity}
\sigma^{\mu\nu} \gamma^5 = \frac{i}{2} \epsilon^{\mu\nu\alpha\beta} \sigma_{\alpha\beta} \; ,
\end{equation}
the axial tensor $\sigma^{\mu\nu} \gamma^5$ is related to 
the tensor $\sigma^{\alpha\beta}$ and thus should not be regarded as 
an independent set for Dirac fermionic DM.
However, for Majorana fermionic $\chi$, such axial tensor structure can 
be present.

Next set of operators are associated with (pseudo) scalar-type exchange
\begin{equation}
{\cal L}_{i=7-10} = \frac{C m_q }{\Lambda_i^3} \,
   \left ( \overline{\chi} \Gamma \chi \right )\;
           \left ( \bar{q} \Gamma^\prime q \right ) \;,
\end{equation}
where $\Gamma , \Gamma^\prime = 1$ or $i\gamma^5$.  The $m_q$ dependence in the
coupling strength is 
included for scalar-type interactions in accord with the trace anomaly in QCD.
We use the current quark masses in the Lagrangian given by \cite{pdg}:
\begin{eqnarray}
 m_u & = & 0.0025\; {\rm GeV}, \quad  m_d \; = \; 0.005 \;{\rm GeV}, \quad
 m_s \; = \; 0.101 \;{\rm GeV}, \nonumber \\
 m_c & = &1.27 \;{\rm GeV}, \quad
 m_b \; = \; 4.19 \; {\rm GeV}, \quad 
 m_t \; = \; 172 \; {\rm GeV}.
\nonumber
\end{eqnarray}
Another light degree of freedom that couples to the Dirac dark matter 
is the gluon field 
\footnote
{We do not study the other gauge bosons, like $W$ and $Z$ bosons, because
they decay into light quarks which then fragment into photons, 
would be softer in this case.}
\begin{eqnarray}
{\cal L}_{i=11-12} &=& \frac{C \alpha_s(2 m_\chi) }{ 4 \Lambda_i^3} \,
   \left ( \overline{\chi} \Gamma \chi \right )\;
          G^{a\mu\nu} G^a_{\mu \nu} \\
{\cal L}_{i=13-14} &=& \frac{C \alpha_s(2 m_\chi) }{4 \Lambda_i^3} \,
   \left ( \overline{\chi} \Gamma \chi \right )\;
          G^{a\mu\nu} \tilde{G}^a_{\mu \nu} 
\end{eqnarray}
where $\Gamma = 1$ or $i\gamma^5$. 
For operators involving gluons, 
the factor of strong coupling constant
$\alpha_s(2 m_\chi)$ is also included in accord with the trace anomaly and
is evaluated at the scale $2m_\chi$ where $m_\chi$ is the dark matter mass.

Finally, we also write down the corresponding operators for complex
scalar dark matter.  Again, we note that the interactions for real scalar
dark matter is similar to complex one and differ by a factor of two. 
We  simply focus on the complex scalar dark matter.  The operators 
corresponding to vector boson exchange are
\begin{equation}
{\cal L}_{i=15,16} =  \frac{C}{\Lambda_i^2} \,
   \left ( \chi^\dagger \overleftrightarrow{\partial_\mu} \chi \right )\;
           \left ( \bar{q} \gamma^\mu \Gamma q \right ) \;,
\end{equation}
where $\Gamma = 1$ or $\gamma^5$ and 
$\chi^\dagger \overleftrightarrow{\partial_\mu} \chi =
 \chi^\dagger (\partial_\mu \chi) - (\partial_\mu \chi^\dagger) \chi$.  
Those corresponding to a scalar boson exchange are
\begin{equation}
{\cal L}_{i=17,18} =  \frac{C m_q }{\Lambda_i^2} \,
   \left (  \chi^\dagger  \chi \right )\;
           \left ( \bar{q} \Gamma q \right ) \;,
\end{equation}
where $\Gamma = 1$ or $i\gamma^5$.  
The corresponding gluonic operators are
\begin{eqnarray}
{\cal L}_{i=19} &=& \frac{C \alpha_s(2 m_\chi) }{4 \Lambda_i^3} \,
   \left ( \chi^\dagger \chi \right )\;
        G^{a \mu\nu} G_{a \mu \nu} \; , \\
{\cal L}_{i=20} &=& \frac{i C \alpha_s(2 m_\chi) }{4 \Lambda_i^3} \,
   \left ( \chi^\dagger \chi \right )\;
        G^{a\mu\nu} \tilde{G}_{a\mu \nu} \; . 
\end{eqnarray}
The whole set of operators we are studying are tabulated in Table~\ref{table1}
together with their corresponding coefficients 
and scaling with velocities in the annihilation cross sections.
Without a particular model in mind we will treat each interaction
independently in our analysis by considering one operator at a time and 
setting the coefficient $C=1$ for simplicity.
\footnote{For gluonic operators, their coefficients
$C$s are induced at loop level and can be smaller.}
The relative importance of each operator can be understood by considering
the non-relativistic expansion of the operator and studying the velocity
dependence.  
It was fully discussed in Ref.~\cite{cty} and hence
we only briefly summarize here for convenience. 
In the non-relativistic limit, 
the spinors for the Dirac DM $\chi$ and $\bar \chi$ annihilation are 
$\psi \simeq  \left( \xi , \epsilon \xi \right)^{\rm T}$ and 
$\bar \psi \simeq  ( \epsilon \eta^\dagger , \eta^\dagger ) \gamma^0$      
where $\xi$ and $\eta$ are two-components Pauli spinors
and 
$\epsilon = O(v/c)$.
We can expand $\bar \psi \gamma^\mu \psi$ as
\begin{eqnarray}
  \bar \psi \gamma^0 \psi &\simeq& 2 \epsilon \eta^\dagger  \xi \, \nonumber \\
  \bar \psi \gamma^i \psi &\simeq & (1+\epsilon^2)
                         \eta^\dagger \sigma_i \xi \, \nonumber 
\end{eqnarray}
where the spatial components are not suppressed by $v/c$.  On the other 
hand, $\bar \psi \gamma^\mu \gamma^5\psi$ in the non-relativistic limit are 
\begin{eqnarray}
  \bar \psi \gamma^0\gamma^5 \psi &\simeq& (1+\epsilon^2)
                               \eta^\dagger  \xi \, \nonumber \\
  \bar \psi \gamma^i \gamma^5 \psi &\simeq& 2 \epsilon \eta^\dagger \sigma_i \xi 
  \, \nonumber 
\end{eqnarray}
where the spatial components are now suppressed by $v/c$. 
It is clear that in the non-relativistic limit the time and spatial
components of the vector and axial vector bilinear 
behave very differently.  We can then consider them separately 
when it is contracted with the trace of the light quark leg. 
If we look at the trace of $(\bar q \gamma^\mu q)$ or 
$(\bar q \gamma^\mu \gamma^5 q)$ in the annihilation amplitude, the
time component part after being squared gives a quantity close to
zero, while the spatial component part gives a quantity
in the order of $m_\chi^2$.  Therefore, it is clear now that 
$\bar \psi \gamma^\mu \psi$ multiplied to $(\bar q \gamma_\mu q)$ or 
$(\bar q \gamma_\mu \gamma^5 q)$ will not be suppressed, while
$\bar \psi \gamma^\mu \gamma^5\psi$ multiplied to $(\bar q \gamma_\mu q)$ or 
$(\bar q \gamma_\mu \gamma^5 q)$ will always be suppressed. 
Therefore, the operators $O_1$ and $O_3$ can contribute to annihilation
much more than the operators $O_2$ and $O_4$. Thus, the limits on 
$O_1$ and $O_3$ are much stronger than $O_2$ and $O_4$.
All the other operators listed in Table~\ref{table1} can be understood similarly \cite{cty}.
We note that some of the operators are doubly suppressed by the velocity of the dark matter
combined with either a light quark mass or strong coupling constant.

In the calculation of the gamma-ray flux from dark matter 
annihilation presented in the next section, 
we only include the light-quark flavors.  We ignore the 
$\chi \overline{\chi} \to t \bar t$ contribution, 
because the $t$ and $\bar t$ first decay into $b W \to b
q \bar q'$ before each light quark undergoes fragmentation into 
hadrons, including pions.
Therefore, the gamma-ray spectrum would be significantly softer than the
direct fragmentation as in $\chi \overline{\chi} \to q \bar q$ \cite{cty}. 

Note that some of the lower limits that we obtain in
Table~\ref{limit} are relative low compared to the dark matter mass.
Notably for the operators $O_{7,9}$, $O_{11,13}$, $O_{15,16}$ and
$O_{17,18}$.  In such cases, one may question the validity of the
effective interaction approach.  The physics behind is easy to
understand. The effects of such operators are very suppressed
because of the small velocity suppression or helicity suppression,
not because of the size of the $\Lambda$.  Therefore, the $\Lambda$
has to be small enough in order to see an effect from these operators.
We argue that the effective momentum
transfer of such velocity-suppressed operators should be 
$m_{\chi} (v/c)$. With $(v/c) \sim 10^{-3}$ for the DM velocity 
at the present epoch, 
as long as the ratio $ m_{\chi} (v/c) / \Lambda$ remains small, 
we expect the effective interaction approach can still be valid.


\section{Gamma-ray Flux}
\subsection{Background Diffuse Gamma Rays}

The Galactic diffuse gamma rays originate primarily from the interactions
of high energy charged particles contained in cosmic rays with the nuclei
in the interstellar medium and the associated radiation fields of the 
charged particles, via a few of the following mechanisms.
\begin{itemize}
\item[{\rm (i)}]  
Gamma-rays coming from the $\pi^0$ decay, which in turn comes from
the interactions of the cosmic rays with the nucleons in the interstellar
medium.  This dominates the background flux for energy higher than 1 GeV.
\item[{\rm (ii)}] 
Inverse Compton scattering occurs when high energy $e^\pm$ collide
with the photons of the interstellar medium.
\item[{\rm (iii)}] 
Bremsstrahlung photons occur when high energy $e^\pm$ are deflected
by the Coulomb field of the interstellar medium.
\item[{\rm (iv)}]
Synchrotron radiation occurs when high energy $e^\pm$ are deflected
by Galactic magnetic field.
\item[{\rm (v)}] 
An extragalactic background, which is expected to be isotropic
and receives contributions from many sources including 
unresolved point sources, diffuse emission from large scale structure formation 
and from interactions between ultra-high energy cosmic rays and relic photons, 
etc.
This background is the least determined and so a fairly large uncertainty
is associated with it. The Fermi-LAT has a measurement of diffuse gamma-ray
in the mid-latitude region and fitted the extra-galactic background by
\begin{equation}
\label{fit}
  E^2 \frac{ d \Phi}{d E} = A \left( \frac{E}{0.281 \; {\rm GeV}} \right)^\delta \;,
\end{equation}
where $A$ and $\delta$ are fitted parameters (the power-law index is 
$\gamma=|\delta - 2|$).
In Ref.~\cite{fermi-lat}, the power-law is fitted to be 
$\gamma = 2.41 \pm 0.05$ 
and $A$ can be determined by the total flux of EGB 
(``extragalactic" diffuse gamma-ray emission)
as $A = ( 0.95\; ^{+0.18}_{-0.17}) \times 10^{-6}
\; {\rm GeV}\,{\rm cm}^{-2}\, {\rm s}^{-1} \, {\rm sr}^{-1}$ for $E > 100$ MeV.
\end{itemize}

\begin{figure}[th!]
\includegraphics[angle=-90,width=5in]{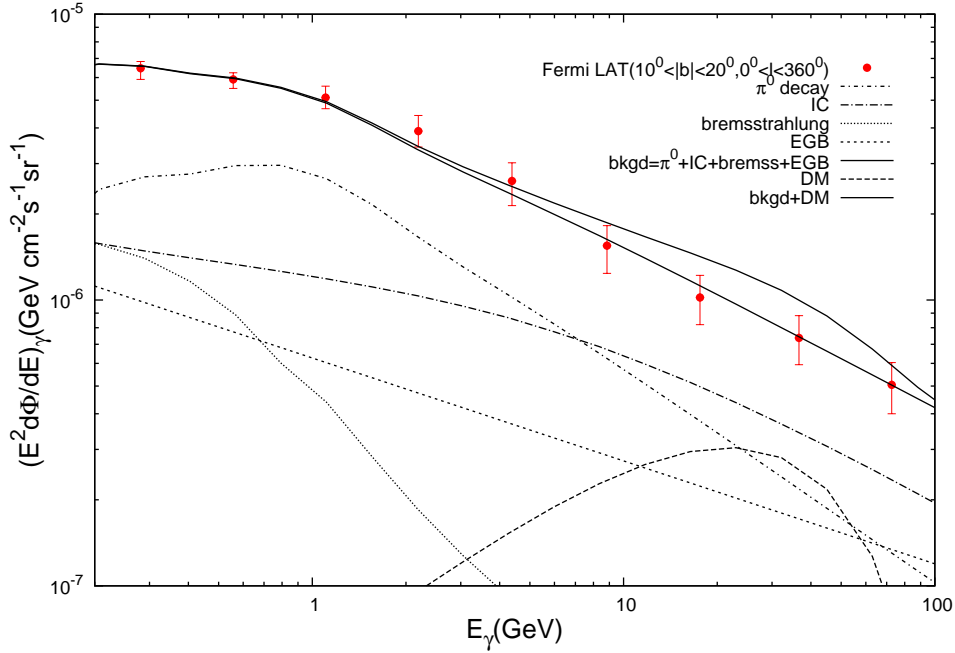}
\caption{\small \label{fig200} 
  The photon spectrum $E^2 (d\Phi/d E_\gamma)$ versus the photon
  energy for the diffuse background gamma rays, including $\pi^0$
  decays (short-dashed-dotted) from interactions of cosmic rays with
  interstellar medium, inverse Compton scattering IC
  (long-dashed-dotted), bremsstrahlung (dotted) and extragalactic EGB
  (short-dashed).  Their sum is shown as the lower solid line.
  A dark matter component due to DM annihilation of $\chi \bar \chi \to 
  q \bar q$ is also shown (long-dashed) and added to the total background 
  (upper solid line).  The DM annihilation is due to a 200 GeV DM particle
  with the effective interaction operator $O_1$ and $\Lambda = 1.5$ TeV.
}
\end{figure}

Since Fermi-LAT did not give details about the parameters of GALPROP
(Cosmic ray propagation code) \cite{GALPROP} 
that they used in their treatment, we run GALPROP 
(the web version) to obtain the various diffuse Galactic backgrounds 
(i) to (iv) and fit EGB component in Eq. (\ref{fit}) to Fermi-LAT
data.  The relevant GALPROP parameters that we used are shown in 
Table~\ref{galprop-para}.  Our fitted EGB is given by
\begin{equation}
    E^2 \frac{ d \Phi}{d E} = (0.99\times 10^{-6}) \, 
  \left(  \frac{E}{0.281 \;{\rm GeV}} \right )^{-0.36} 
\;\; \mbox{GeV cm$^{-2}$ s$^{-1}$ sr$^{-1}$ }
\end{equation}
which gives a power-law index $\gamma = 2.36$ and $E$ is in GeV.
It is close enough
to the one obtained by Fermi-LAT.  The $\chi^2 = 2.435$ for 7 d.o.f. 
The various curves are all within the uncertainties quoted in the Fig.~6(a)
of the Fermi-LAT paper \cite{fermi-lat}.  Various diffuse background curves 
and their sum are shown in Fig.~\ref{fig200}.

\begin{table}[b!]
\caption{\small \label{galprop-para}
The values of the parameters that we used in running the code
GALPROP to reproduce the background curves in 
Figs.~\ref{fig200}--\ref{fig500}. 
For unstated parameters we employ the default values.}
\small
\begin{ruledtabular}
\begin{tabular}{lr}
Parameter (unit)       &   Value \\
\hline
Minimum galactocentric radius $r_{min}$ (kpc) &
 $00.0$ \\
Maximum galactocentric radius $r_{max}$ (kpc) &
 $25.0$ \\
Minimum height $z_{min}$ (kpc) &
 $-04.0$ \\
Maximum height $z_{max}$ (kpc) &
 $+04.0$ \\
ISRF factors for IC calculation: optical, FIR, CMB $ISRF_{factors}$ & 
 $1.9,1.9,1.9$ \\
Gamma-ray Intensity Skymap Longitude minimum $long_{min}$ (degrees) &
 $0.0$ \\
Gamma-ray Intensity Skymap Longitude maximum $long_{max}$ (degrees) &
 $360.0$ \\
Gamma-ray Intensity Skymap Latitude minimum $lat_{min}$ (degrees) &
 $+10.0$ \\
Gamma-ray Intensity Skymap Latitude maximum $lat_{max}$ (degrees) &
 $+20.0$ \\
Binsize in Longitude for Gamma-ray Intensity Skymaps $d_{long}$ (degrees) &
 $0.50$ \\
Binsize in Latitude for Gamma-ray Intensity Skymaps $d_{lat}$ (degrees) &
 $0.50$ \\
Diffusion Coefficient Normalization $D_{0xx}$
   ($10^{28}\, {\rm cm}^2 {\rm s}^{-1}$) &      $6.1$  \\
Diffusion Coefficient Index Below Break Rigidity $D_{g1}$ &
       $0.33$ \\
Diffusion Coefficient Index Above Break Rigidity $D_{g2}$ &
       $0.33$ \\
Diffusion Coefficient Break Rigidity $D_{{rigid}\; {br}}$ ($10^3\,{\rm  MV}$) &
 $4.0$  \\
Alfven Speed $v_A$ (km s$^{-1}$) &
       $30$ \\
Nuclear Break Rigidity $nuc_{{rigid}\; {br}}$ ($10^3$~MV) &
 $10.0$ \\
Nucleus Injection Index Below Break Rigidity $nuc_{g1}$ &
 $2.00$ \\
Nucleus Injection Index Above Break Rigidity $nuc_{g2}$ &
 $2.43$ \\
Proton Flux Normalization
($10^{-9}\,{\rm cm}^{-2}\,{\rm sr}^{-1}\, {\rm s}^{-1}\, {\rm MeV}^{-1}$) &
 $4.90$ \\
Proton Kinetic Energy for Normalization ($10^5$~MeV) &
 $1.00$ \\
Electron Break Rigidity0 $electron_{{rigid}\; {br0}}$ ($10^4$~MV) &
 $3.0$ \\
Electron Break Rigidity $electron_{{rigid}\; {br}}$ ($10^9$~MV) &
 $1.0$ \\
Electron Injection Index Below Break Rigidity0 $electron_{g0}$ &
 $2.20$ \\
Electron Injection Index Above Break Rigidity0 and Below Break Rigidity $electron_{g1}$ &
 $2.54$ \\
Electron Injection Index Above Break Rigidity $electron_{g2}$ &
 $2.5$ \\
Electron Flux Normalization
($10^{-10}\,{\rm cm}^{-2}\,{\rm sr}^{-1}\, {\rm s}^{-1}\, {\rm MeV}^{-1}$) &
 $4.0$ \\
Electron Kinetic Energy for Normalization ($10^4$~MeV) &
 $3.45$ 
\end{tabular}
\end{ruledtabular}
\end{table}

\begin{figure}[th!]
\includegraphics[angle=-90,width=5in]{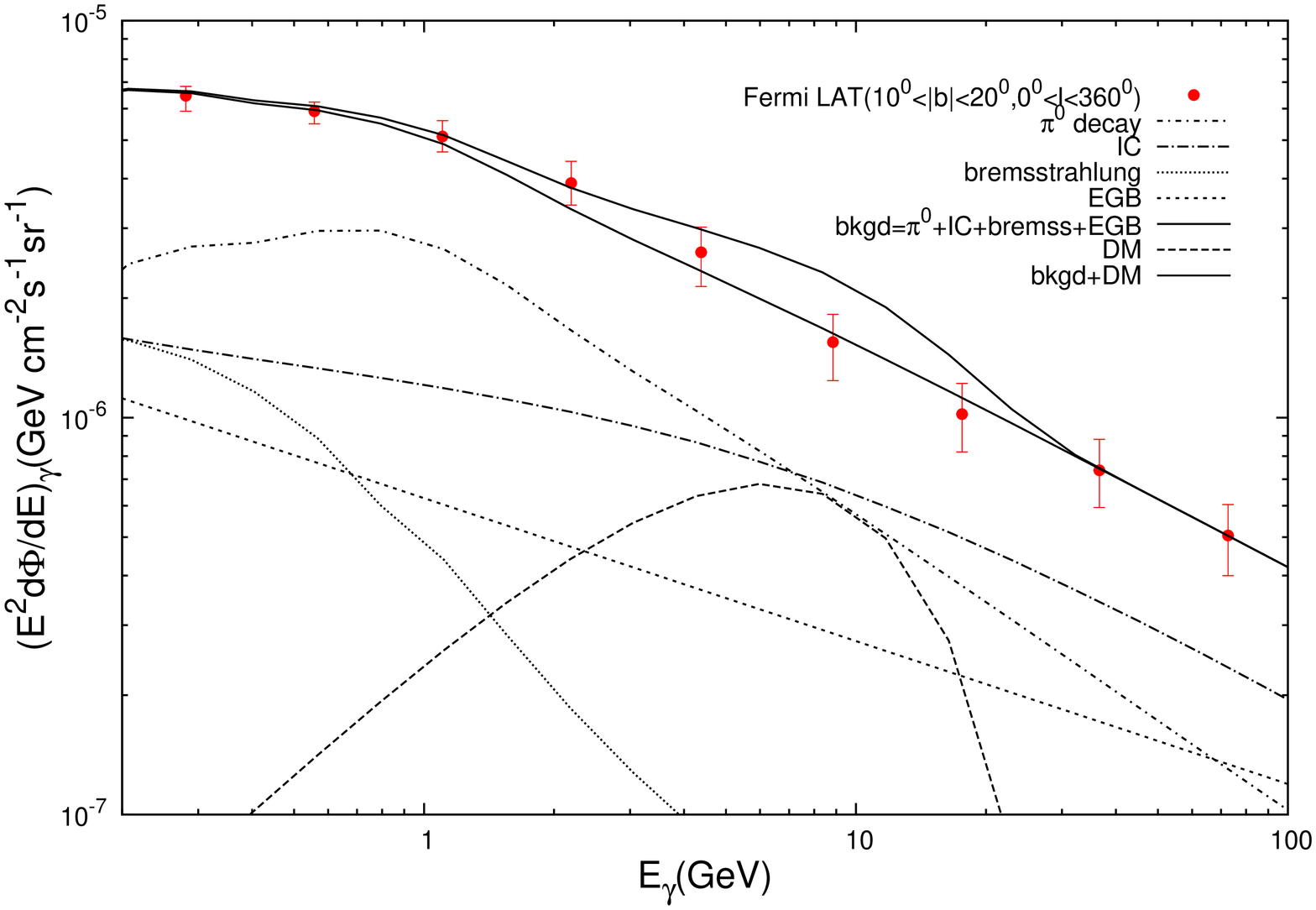}
\caption{\small \label{fig50} 
 Same as Fig.~\ref{fig200} but the DM annihilation is due to a 50 GeV dark
 matter particle with the effective interaction operator $O_1$ and
 $\Lambda =0.87$ TeV.
}
\end{figure}

\begin{figure}[th!]
\includegraphics[angle=-90,width=5in]{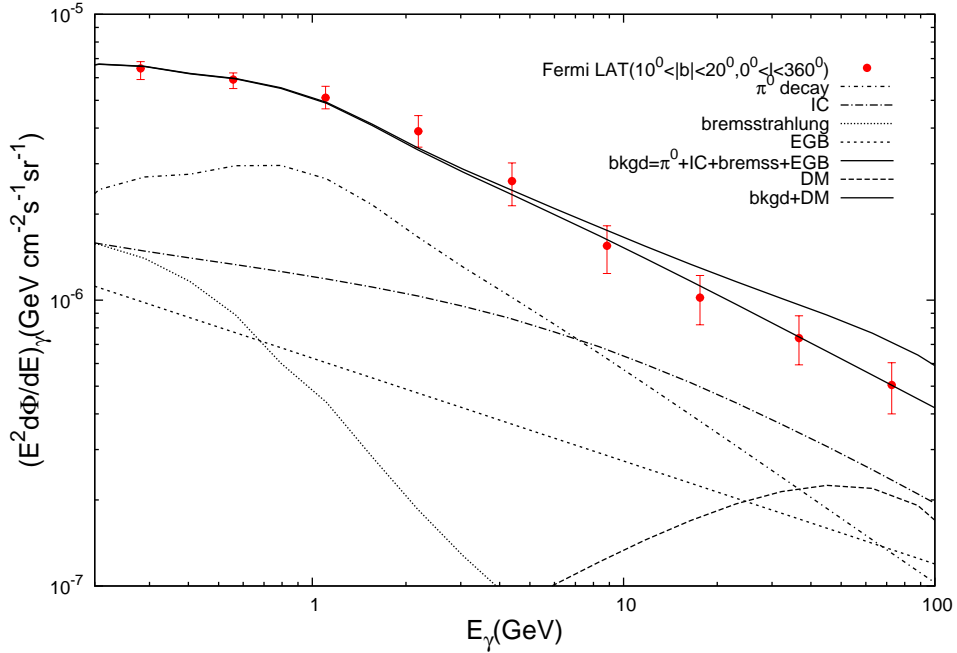}
\caption{\small \label{fig500} 
 Same as Fig.~\ref{fig200} but the DM annihilation is due to a 500 GeV dark
 matter particle with the effective interaction operator $O_1$ and
 $\Lambda =1.9$ TeV.
}
\end{figure}

\subsection{Dark Matter Annihilation}

The dominant DM contribution to photon flux in this scenario comes from
\begin{equation}
\chi \overline{\chi} \to q \bar q \to \pi^0 + X \to 2 \gamma + X \;,
\end{equation}
in which all the $q, \bar q\; (q=u,d,c,s,b)$ have probabilities 
fragmenting into $\pi^0$, which then decay almost entirely into 
two photons.
As mentioned in the Introduction, 
we employ two approaches of obtaining the photon spectrum due to
fragmentation of light quarks. (i) Using the process 
$e^+ e^- \to q \bar q$ with initial radiations turned off in 
Pythia \cite{pythia} and extracting the photon spectrum in the final state. 
The photon mainly comes from the decay of $\pi^0$, which are in turn 
produced by fragmentation of light quarks, plus 
a very small fraction from the bremsstrahlung photon off the quark legs.  
(ii) Using the fragmentation function of
$q,\bar q, g$ into $\pi^0$ from the fitting of Ref.~\cite{kniehl}, then
convoluting with the $dN/dE_\gamma (\pi^0 \to \gamma \gamma)$  to obtain
the photon spectrum of quarks into photon. 
Both approaches give the photon spectra close enough to each other 
for our purpose of numerical calculations.
Thus, the resulting limits using both approaches 
are also the same within numerical accuracy.
However, we have to use the second approach when we place limits on the
operators involving gluons in the annihilation, because we do not 
find a simple process in Pythia for extraction of $g\to \gamma$ 
fragmentation.

The photon flux is proportional to the square of the number density of the
DM particles $(\rho/m_\chi)^2$, the annihilation cross section 
$\langle \sigma v \rangle$ and the spectrum of photons $d N/dE_\gamma$ per 
annihilation.  The flux observed is found by integrating the number 
density squared along the line-of-sight 
connecting from the source to the observer, given by
\begin{equation}
  \Phi = \frac{\langle \sigma v \rangle}{2} \, \frac{dN}{dE_\gamma} \,
  \frac{1}{4 \pi m_\chi^2}\, \int_{\mbox{\scriptsize line of sight}}\;
 ds \, \rho^2 (s, \psi) \;, 
\end{equation}
where $s$ runs along the line of sight and $\psi$ is the angle from
the direction of the Galactic Center.  Here the factor of $2$ in the
denominator accounts for particle-antiparticle annihilation 
since we 
are dealing with Dirac or complex scalar DM. We employ the
isothermal profile in running the GALPROP \cite{GALPROP}.

The contribution from annihilation of DM to the observed diffuse photon 
spectrum is shown in each Fig.~\ref{fig200} -- Fig.~\ref{fig500} (the
long dashed curve near the bottom of each figure.)  The position of
the peak of the DM curve corresponds to about $0.1 m_\chi$, 
as can be clearly spotted at each of these figures.
When we add the DM contribution to the total background curve, we 
see the deviation from the data points. We can quantify the deviation
from the data by calculating the total $\chi^2$ as a function of 
DM mass and the interaction strength of the annihilation (given by
$\Lambda^2$).

Here we adopt a simple statistical measure to quantify the effect
of each operator.  We calculate the $3\sigma$ limit on each scale
$\Lambda_i$. We assume the data agree well with the expected background,
and then we calculate the $\chi^2$ with finite $\Lambda_i$'s
until we obtain a $\chi^2$ difference of 
$\Delta \chi^2 \equiv \chi^2 - \chi^2_{\rm bkgd} = 9$  ($3\sigma$).  
The DM curve in each figure corresponds to a $3\sigma$ deviation from the
data points.  We calculate the limit for each operator and list them
in Table~\ref{limit}.
For those unsuppressed operators the limit is of
order 1 TeV for $m_\chi = 50 -500$ GeV. 
But for those operators
suppressed by the velocity of the DM, light quark masses or strong coupling
constant, the limit is significantly weaker of order $0.01 - 0.1$ TeV.

\begin{figure}[t!]
\includegraphics[width=6in]{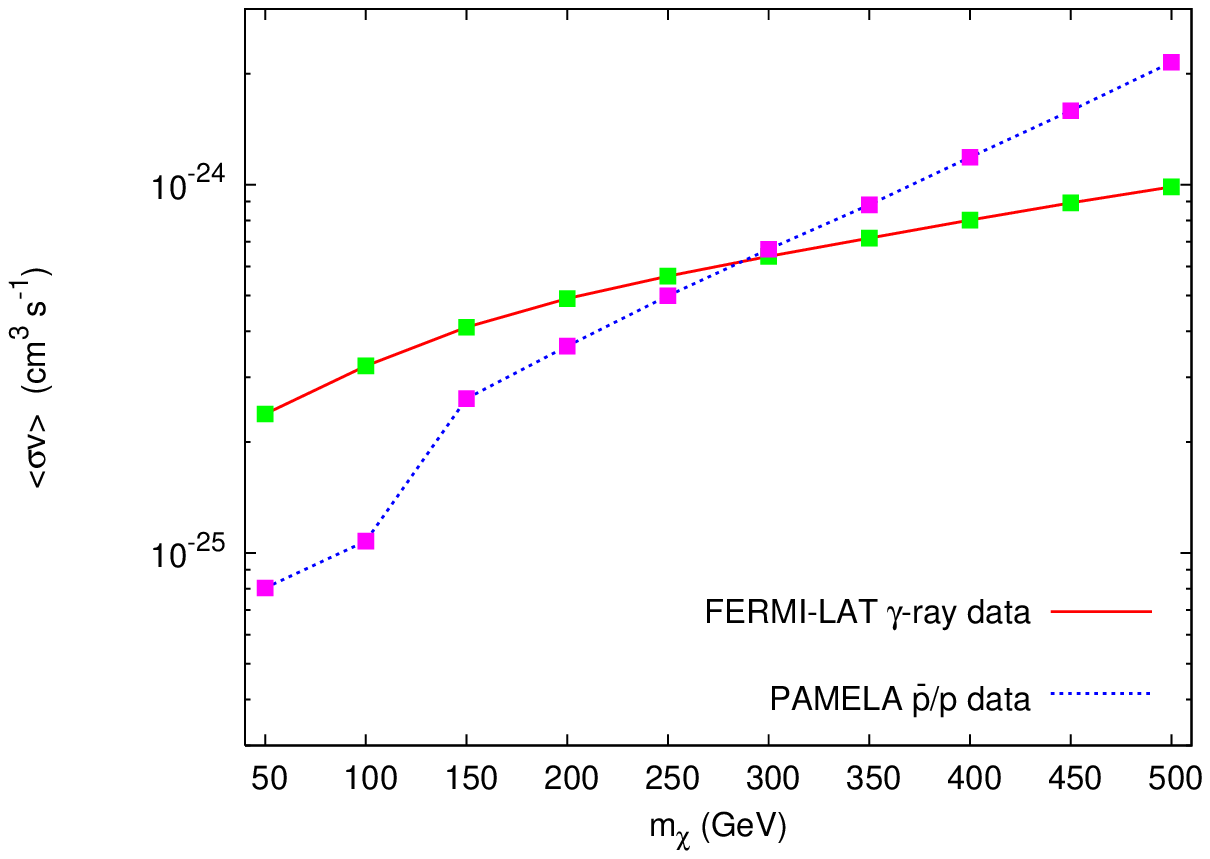}
\caption{\small \label{v_sigma}
The $3\sigma$ upper limits on the annihilation cross section
$\sigma v(\chi \bar \chi \to q \bar q)$ versus the DM mass
due to the Fermi-LAT photon-flux data 
($10^\circ < |b| < 20^\circ,\; 0^\circ < l < 360^\circ$)
and due to the PAMELA antiproton-flux data \cite{cty}.  
The limits are approximately
independent of the operators. The annihilation cross sections above the
curves are ruled out. The $x$-axis is at the value $3\times 10^{-26}\,
{\rm cm}^{3}\, {\rm s}^{-1}$ which is the required annihilation cross
section to give the correct thermal relic density.
}
\end{figure}

Once we obtained the lower limits on $\Lambda$ for each DM mass, we
can calculate the corresponding upper limits on 
$\langle \sigma v \rangle (\chi \bar \chi \to q \bar q)$ 
as a function of DM mass. 
We found that these upper limits on $\langle \sigma v \rangle
( \chi \bar \chi \to q \bar q)$
are approximately independent of the operators with $q$ and $\bar q$ 
in the final state.  We understand this as the fragmentation process of
$q$ or $\bar q$ into $\pi^0$ followed by $\pi^0 \to \gamma\gamma$ decays
does not depend on how the quarks are produced.  As long as the quarks
produced have the same energy, the fragmentation rate or pattern should be
the same. We show in Fig.~\ref{v_sigma} the upper limits on
$\langle \sigma v \rangle ( \chi \bar \chi \to q \bar q)$ 
allowed by the Fermi-LAT
photon-flux data, as well as those obtained in Ref.~\cite{cty}
using the antiproton-flux data of PAMELA \cite{pamela-p}.
It is clear in Fig.~\ref{v_sigma} that the antiproton-flux data from
PAMELA does give a stronger constraint on the annihilation cross section
 $\langle \sigma v \rangle (\chi \bar \chi \to q \bar q )$ 
than the Fermi-LAT photon-flux
data for lighter DM ($50 - 300$ GeV), while Fermi-LAT gamma-ray data
does constrain stronger for heavier DM mass ($300-500$ GeV).

\begin{table}[th!]
\caption{\small \label{limit}
The $3\sigma$ lower limits on the effective sale $\Lambda$ of each 
operator listed in Table~\ref{table1}.
We take the coefficient $C=1$ with $m_\chi$ = 50, 100, 200 and 500 GeV.
}
\smallskip
\begin{ruledtabular}
\begin{tabular}{lrrrr}
Operators       &   \multicolumn{4}{c}{$\Lambda$ (TeV)} \\
                & $m_\chi$  (GeV)  = 50  & 100   &  200   & 500 \\
\hline
\multicolumn{4}{l}{Dirac DM, (axial) vector/tensor exchange} \\
\hline
$O_1 = (\overline{\chi} \gamma^\mu \chi)\, (\bar q \gamma_\mu q)$ &
      $0.87$ & $1.15 $ & $1.46$ & $1.94$  \\
$O_2 = (\overline{\chi} \gamma^\mu \gamma^5\chi)\, (\bar q \gamma_\mu  q)$ &
       $0.025$ & $0.033 $ & $0.042$ & $0.055$ \\
$O_3 = (\overline{\chi} \gamma^\mu \chi)\, (\bar q \gamma_\mu \gamma^5 q)$ &
     $0.87$  &  $1.15 $ & $1.46$ & $1.94$ \\
$O_4 = (\overline{\chi} \gamma^\mu \gamma^5 \chi)\, 
    (\bar q \gamma_\mu \gamma^5 q)$ &  
 $0.13$ & $0.12$ & $0.11$ & $0.10$  \\
$O_5 = (\overline{\chi} \sigma^{\mu\nu} \chi)\, (\bar q \sigma_{\mu\nu} q)$ & 
       $1.04$  &  $1.36 $  &  $ 1.74 $  & $2.31$ \\
$O_6 = (\overline{\chi} \sigma^{\mu\nu} \gamma^5 \chi)\, 
  (\bar q \sigma_{\mu\nu} q)$ & $1.04$ & $1.36 $ &  $1.74 $ & $2.31$\\
\hline
\multicolumn{4}{l}{Dirac DM, (pseudo) scalar exchange} \\
\hline
$O_7 = (\overline{\chi}  \chi)\, (\bar q  q)$ & $0.009$ 
                        & $0.011 $ & $0.014 $ & $0.017$\\
$O_8 = (\overline{\chi} \gamma^5  \chi)\, (\bar q  q)$ & 
             $0.094$&    $0.11 $ &  $0.14$ & $0.17$ \\
$O_9 = (\overline{\chi}  \chi)\, (\bar q \gamma^5 q)$&
    $0.009$ & $0.011 $ &    $ 0.014$ & $0.017$ \\
$O_{10} = (\overline{\chi} \gamma^5 \chi)\, (\bar q \gamma^5 q) $ &  
   $0.094$ &  $0.11 $  & $0.14$ & $0.17$ \\
\hline
\multicolumn{4}{l}{Dirac DM, gluonic} \\
\hline
$O_{11} = (\overline{\chi} \chi)\, G_{\mu\nu} G^{\mu\nu}$ & 
     $0.011$ &  $0.013 $  &  $ 0.017$  & $0.024$ \\
$O_{12} = (\overline{\chi} \gamma^5 \chi)\, G_{\mu\nu} G^{\mu\nu} $ & 
     $0.11$ &   $0.13 $   &     $ 0.17$  & $0.24$ \\
$O_{13} = (\overline{\chi} \chi)\, G_{\mu\nu} \tilde{G}^{\mu\nu}$ & 
     $0.011$ &   $0.013 $  &  $ 0.017 $ & $0.024$ \\
$O_{14} = (\overline{\chi} \gamma^5 \chi)\, G_{\mu\nu} \tilde{G}^{\mu\nu} $& 
     $0.11$ &  $0.13$  &  $ 0.17  $ & $0.24$\\
\hline
\hline
\multicolumn{4}{l}{Complex Scalar DM, (axial) vector exchange} \\
\hline
$O_{15} = (\chi^\dagger \overleftrightarrow{\partial_\mu} \chi)\, 
  ( \bar q \gamma^\mu q ) $&
    $0.025$ &    $0.033 $ &  $0.042$ & $0.055$ \\
$O_{16} = (\chi^\dagger \overleftrightarrow{\partial_\mu} \chi)\, 
( \bar q \gamma^\mu \gamma^5 q ) $&
    $0.025$ &  $0.033$ & $ 0.042  $ & $0.055$ \\
\hline
\multicolumn{4}{l}{Complex Scalar DM, (pseudo) scalar exchange} \\
\hline
$O_{17} = (\chi^\dagger \chi)\, ( \bar q  q ) $&
    $0.11$ &     $0.10 $   &        $ 0.095 $ & $0.083$ \\
$O_{18} = (\chi^\dagger \chi)\, ( \bar q  \gamma^5 q ) $&
    $0.11$ &     $0.10 $     &  $ 0.095   $ & $0.083$ \\
\hline
\multicolumn{4}{l}{Complex Scalar DM, gluonic} \\
\hline
$O_{19} = (\chi^\dagger \chi)\, G_{\mu\nu} {G}^{\mu\nu} $   & 
     $0.13$ &       $0.13$ &  $ 0.13  $ & $0.14$ \\
$O_{20} = (\chi^\dagger \chi)\, G_{\mu\nu} \tilde{G}^{\mu\nu} $& 
     $0.13$ &      $0.13$      &     $  0.13  $& $0.14$ \\
\end{tabular}
\end{ruledtabular}
\end{table}

\section{Discussion and Conclusions}

Here we do a comparison with the limits obtained in collider 
\cite{cao,bai,tait}, gamma-ray lines
in Ref.~\cite{tait-gamma}, antiproton flux \cite{cty} 
and direct searches.

\begin{itemize}

\item {\it Comparison to limits obtained using gamma-ray lines
    {\rm\cite{tait-gamma}} against the Fermi-LAT data} \cite{fermi-lat}.
  As mentioned before, our continuum gamma-ray signals are obtained via quark
  or antiquark fragmentation, whereas in Ref.~\cite{tait-gamma} the discrete gamma-ray line is
  coming from one-loop dressing of the effective operators.
  Note also that we are calculating the $3\sigma$ lower limit while
  Ref.~\cite{tait-gamma} reported the 95\% C.L. lower limits, which
  are approximately $2 \sigma$, so our limits are slightly more
  conservative.  
  We found that the limits for operator $O_{1,3}$ are
  substantially better than the corresponding operators $D_{5,7}$ of
  Ref.~\cite{tait-gamma}.  Our limits are $0.9 - 1.9$ TeV for
  $m_{\chi} = 50 - 500$ GeV, while their limits are $0.12 - 0.6$ TeV
  for $m_{\chi} = 50 - 200$ GeV.  We see in this case (Dirac DM with
  vector-boson exchange), the contribution to photon flux through the
  continuum ($q \to \pi^0 \to \gamma$) is substantially better than
  the discrete line spectrum (via the loop process) when compared
  against the Fermi-LAT data.  The limits for operators $O_{7-10}$
  (Dirac DM with scalar-boson exchange) and $O_{17,18}$ (complex
  scalar DM with scalar-boson exchange) are about the same as
  $D_{1-4}$ and $C_{1,2}$, respectively, of Ref.~\cite{tait-gamma}. On
  the other hand, limits for operators $O_{15,16}$ (complex scalar DM with
  vector-boson exchange) are weaker than $C_{3,4}$ of
  Ref.~\cite{tait-gamma}.

\item {\it Comparison to limits obtained using antiproton flux
    {\rm\cite{cty}} against the PAMELA antiproton data} \cite{pamela-p}.
  The approach that we did in Ref.~\cite{cty} is very similar to the
  work here, except that we require to see photon instead of
  antiproton in the final state and we are comparing with two
  entirely different categories of data.  We found that at the lower
  end $m_\chi \sim 50 -300 $ GeV the limits from antiproton data are
  somewhat better than those from photon-flux data.  This is because
  of the nature of the fragmentation and decay chain: one fragments
  into antiproton directly while the other one fragments into $\pi^0$
  then followed by decays of $\pi^0\to \gamma\gamma$. The peak of the
  antiproton spectrum occurs at larger energies than the peak of the
  photon spectrum for the same DM mass.  On the other hand, when
  $m_\chi$ increases to about $300-500$ GeV, limits from gamma-ray flux
  constrains stronger on the DM interactions.

\item 
  {\it Comparison to limits obtained using collider data}
   \cite{bai,tait}.  The collider limits on $\Lambda$ come from
  monojet plus missing energy data. Basically, for operators that are
  velocity suppressed, e.g., $O_{2,4}$ among $O_{1-4}$, the limits
  from photon flux are not as good as those from colliders; and vice
  versa for operators that are not velocity suppressed, e.g.,
  $O_{1,3}$ among $O_{1-4}$, The collider limits for $O_{1-4}$ are
  $200 - 300 $ GeV, while the limits from photon flux are $0.9 - 1.9$
  TeV for $O_{1,3}$ but only $25 - 100$ GeV for $O_{2,4}$. 
  The operators for which photon flux gives
  better limits are $O_{1,3}$, $O_{5,6}$, $O_{8,10}$, $O_{12,14}$,
  $O_{17,18}$ and $O_{19,20}$.

\item 
  {\it Comparison to limits obtained by the direct detection experiments} \cite{cdms,xenon}.
Two most stringent experiments on the spin-independent (SI) cross sections come
from CDMS \cite{cdms} and XENON100 \cite{xenon}.  The best upper limit 
from CDMS \cite{cdms} is 
$\sigma^{\rm SI}_{\chi N} \simeq 3.8 \times 10^{-44} \;{\rm cm}^2$ 
at $m_\chi =70$ GeV and from XENON100 \cite{xenon} is 
$\sigma^{\rm SI}_{\chi N} \simeq 0.7 \times 10^{-44} $ at $m_\chi =50$ GeV.
Note that only the operators $O_1$ and $O_7$ for Dirac DM
contribute to spin-independent cross sections.  For $O_7 = m_q
(\bar \chi \chi)(\bar q q)/\Lambda_7^3$ the SI cross section  is given by
\begin{equation}
 \sigma^{\rm SI}_{\chi N} = \frac{\mu_{\chi N}^2}{\pi} | G^N_s|^2 \;,
\end{equation}
where 
$\mu_{\chi N}$ is the reduced mass of the DM and nucleon, and
\begin{equation}
  G^N_s = \sum_{q} \langle N | \bar q q| N \rangle \left( \frac{m_q}{\Lambda^3}
  \right ) \;, \quad 
  \langle N | \bar q q| N \rangle = \frac{m_N}{m_q} \times
  \left\{ \begin{array}{r@{\quad:\quad}l}
  f^N_{Tq}   & {\rm light \, quarks} \\
  \frac{2}{27} f^N_{Tg}  & {\rm heavy \, quarks}
  \end{array}
  \right.\;.
\end{equation}
Using the default values of $f^N_{Tq}$ and $f^N_{Tg}$ adopted in DarkSUSY \cite{darksusy}
and averaging between the neutron and proton for $\sigma^{\rm SI}_{\chi N}$, we obtain
\begin{equation}
 \sigma^{\rm SI}_{\chi N} \simeq \frac{m_N^4}{\pi \Lambda_7^6} (0.3769)^2 \;.
\end{equation}
Using the limit $ \sigma^{\rm SI}_{\chi N} < 10^{-44}\;{\rm cm}^{2}$ for 
$m_\chi \sim 50 - 200$ GeV \cite{xenon},  we obtain 
\begin{equation}
 \Lambda_7 > 330 \;{\rm GeV}  \;.
\end{equation}
This limit is much better than the limit on $O_7$ shown in Table~\ref{limit}.
On the other hand, with the operator $O_1 = (\bar \chi \gamma^\mu \chi)
(\bar q \gamma_\mu q)/\Lambda_1^2$ we obtain
\begin{equation}
 \sigma^{\rm SI}_{\chi N} = \frac{\mu_{\chi N}^2}{256 \pi} | b_N|^2 \;,
\end{equation}
where $ b_N \simeq \frac{3}{2} ( \alpha_u^V + \alpha_d^V)$, which
is obtained by taking the average of neutrons and protons inside a nucleon,
and 
$\alpha_{u,d}^V = 1/\Lambda_1^2$ are the coefficients in front of the operators for the valence
$u$ and $d$ quark.  We therefore obtain
\begin{equation}
 \sigma^{\rm SI}_{\chi N} \simeq \frac{9}{256 \pi} \frac{m_N^2}{\Lambda_1^4}\;,
\end{equation}
which gives the limit on $\Lambda_1$, with 
$ \sigma^{\rm SI}_{\chi N} < 10^{-44}\;{\rm cm}^{2}$, as 
\begin{equation}
 \Lambda_1 > 4.4 \;{\rm TeV} \;.
\end{equation}
It is about $2-3$ times better than the corresponding limits on $O_1$
shown in Table~\ref{limit}.
\end{itemize}

In summary, we have used an effective interaction approach to investigate
the effects of dark matter interactions with light quarks on 
diffuse photon flux coming from the mid-latitude
$(10^\circ < |b| < 20^\circ, \; 0^\circ < l < 360^\circ )$ region.
We have assumed a standard halo density 
while using the GALPROP to calculate the resulting
diffuse gamma-ray spectrum.  The diffuse gamma-ray background includes
$\pi^0$ decays produced by scattering of cosmic rays with the interstellar
medium, inverse Compton scattering, bremsstrahlung, synchrotron radiation
and extragalactic diffuse gamma rays.  The background can reproduce
the data recorded by Fermi-LAT very well.  
Dark matter annihilation provides yet another source of diffuse gamma rays.  
The dominant mechanism is annihilation into light quarks, followed by fragmentation 
into neutral pions, which then further decay into photons.  The gamma-ray
spectrum has no particular feature but a continuum; nevertheless, the flux
is large.  We have shown that the effective interactions of the DM
can give rise to a nontrivial contribution to the diffuse gamma-ray spectrum,
as presented in various figures. 

We have successfully used the data to
obtain $3\sigma$ limits on the scale $\Lambda_i$. The best limits
are from the Dirac DM with vector or tensor boson exchanges.  
The limits for $O_{1,3}$ and $O_{5,6}$ 
are about $1 - 2$ TeV,  while other operators suppressed either 
by the velocity of dark matter, light quark masses or strong coupling constant 
give milder limits.
Note that these limits from photon flux are {\it lower} limits on
$\Lambda_i$.  We found that the limits obtained are very comparable to
those obtained using collider \cite{bai,tait},
gamma-ray line search \cite{tait-gamma} and anti-matter experiments \cite{cty}.
However, the two operators $(\bar \chi \chi)(\bar q q)$
and $(\bar \chi \gamma^\mu \chi)(\bar q \gamma_\mu q)$, which
contribute to spin-independent cross sections, are constrained more
severely by the recent XENON100 data.


\section*{Acknowledgments}
This work was supported in parts by the National Science Council of
Taiwan under Grant Nos. 99-2112-M-007-005-MY3 and
98-2112-M-001-014-MY3 as well as the
WCU program through the KOSEF funded by the MEST (R31-2008-000-10057-0).

\end{document}